\documentclass[epj]{svjour} %{svjour2}

\usepackage{color}
\usepackage{graphics}
\usepackage{amsmath}
\usepackage{amssymb}
\begin{document}
\title{Calculation of survival probabilities for cells exposed to high ion fluences}
%\title{How to make the cell survival probability uniform along a spread-out Bragg peak}
\author{Eugene Surdutovich\inst{1}
    \and Andrey V. Solov'yov\inst{2,3}
}                     % Do not remove
\offprints{Eugene Surdutovich}          % Insert a name or remove this line
\institute{Department of Physics, Oakland University, Rochester, Michigan 48309, USA
     \and
MBN Research Center,
% Altenhoferallee 3
60438 Frankfurt am Main, Germany
\and On leave from A.F. Ioffe Physical Technical Institute, % Politechnicheskaya 26,
194021 St. Petersburg, Russian Federation}
\authorrunning{E. Surdutovich and A.V. Solov'yov}
\titlerunning{Cell survival along SOBP}

\date{Received: date / Revised version: date}
\abstract{A methodology of calculations of survival curves with an account for ion paths interference is developed using the multiscale approach to the physics of radiation damage with ions. The method is applied to different targets and shouldered survival curves are obtained. The recipe is designed for both high and low values of linear energy transfer.
} %end of abstract
\maketitle

\section{Introduction}

Cell survival curves, which give the dependence of probability of cell survival depending on the dose are one of the most important pieces of information for radiation treatment planning. Survival curves for numerous cell lines exposed to x-rays have been studied for decades, while the studies of cells exposed to proton and heavier ion beams are more recent developments. Most of these studies are experimental even though the theoretical models have appeared in 1950s and have substantially developed since then~\cite{Alpen,Katz67,MKModel,BeuveIBCT}. The generic look of these curves in semilogarithmic scale (dose on abscissa and logarithm of sell survival probability on ordinate) is parabola-like with the vertex in the origin, corresponding to 100\% survival at zero dose.

The parabola has been described as
\begin{eqnarray}
\ln \Pi_{surv}=-\alpha d - \beta d^2~,
\label{lqm}
\label{densities}
\end{eqnarray}
where $d$ is the dose and $\alpha$ and $\beta$ are the coefficients. The largeness of ratio $\beta/\alpha$ determines the degree to which the curve is shouldered vs. a straight line when $\beta=0$. Straight lines correspond to ``single hit'' scenarios, i.e., to situations when a single hit of a target with some fixed probability causes lethal damage. For instance, many survival curves for cells exposed to ions are nearly straight lines~\cite{MSAColl,CellSurSR16}. Shouldered lines can be caused by a number of effects and there is a large variety of shouldered curves~\cite{Alpen,Hall}. However, two major ideas of their origin are the effects of biological repair (which is considered to be the major cause) and of interference of tracks. In Ref.~\cite{CellSurSR16}, the biological repair effect on survival curves has been studied using the multiscale approach to the physics of radiation damage with ions (MSA)~\cite{MSAColl,pre,IBCTbook}. This approach has been conceived by the authors of this paper in order to understand the radiation damage with ions on a quantitative level focusing on particular physical, chemical, and biological effects that bring about lethal damage to cells exposed to ion beams.

Cell survival curves for a large variety of cell lines are being studied experimentally in many places around the world. Most of these experiments are done {\em in vitro} in order to predict what may happen to the same cells {\em in vivo}. The applied doses roughly correspond to those, which are going to be used during the treatment. These doses are not very large, which means that the fluences of protons or heavier ions are small enough so that the ion tracks do not overlap with each other. However, in many {\em in vitro} experiments, larger doses are used~\cite{Kowalska17,Fromm,Thomas1,Thomas2}. Then such an overlap is expected. This work is devoted to the development of a method that allows to consider large doses within the MSA.

The approach to this problem is similar to the MSA's treatment of radiation damage when the tracks do not overlap~\cite{MSAColl,CellSurSR16}. The main difference of the MSA is that instead of trying to use the linear-quadratic model (LQM) given by (\ref{lqm}) and assume that the damage is ``done'' by dose, i.e., energy, the actual effects responsible for cell damage are analyzed one by one. These effects are the due to secondary electrons that are produced in the process of ionization of tissue by ions, by reactive species that are produced in ion tracks as a consequence of ionization and excitation of medium as well as possible radiolysis of water molecules in the cores of ion tracks. In order to calculate the cumulative effect of these agents, we analyze their production and transport towards targets distributed in the surrounding of ion paths. In this work, the target is considered to be a cylinder surrounding a single twist of nuclear DNA.

\section{Calculation of lethal lesions yield in the case of two interfering ion's paths}

In this section, we calculate the yield of lethal lesions as a result of irradiation of tissue with a high-dose radiation using the MSA. The key difference of this calculation from other approaches is that a lethal lesion is considered to be a result of several physical, chemical, and biological processes rather than exposure of a given volume to a given dose, i.e., absorption of some energy.

The scenario of radiation damage is as follows. An ion traverses a cell nucleus. It ionizes molecules (mostly water) on its passage. Ejected secondary electrons (first generation) start with average energy of $\sim 45$-eV. They lose most of this energy within 1-1.5~nm of the ion's path ionizing more molecules (including biomolecules). The second generation of slower electrons is formed. These secondary electrons can cause damage only within a tiny region of a few nm. This damage can be estimated from the calculation of average number of secondary electrons incident on uniformly distributed targets (DNA segments) in the region~\cite{MSAColl,pre}. A quantity ${\cal N}_e(r)$, the average number of simple lesions on a target at a distance $r$ from the ion's path is calculated as a result. This is the secondary electrons' contribution and in this work the difference is that the electrons may come not just from a single ion, but from two (or even more) ions. To not overcomplicate this problem, we will consider all ions to have same energies and, therefore, same values of linear energy transfer (LET).

Most of the reactive species (free radicals and solvated electrons) are formed at locations of ionizations described above~\cite{Radicals}. If the LET is relatively small, the number of reactive species is small as well and their interaction can be neglected. Then they very slowly (compared to secondary electrons) diffuse reacting with DNA targets on their way. A quantity ${\cal N}_r(r)$, the average number of simple lesions due to reactive species on a target at a distance $r$ from the ion's path is calculated as a result. This is the reactive species contribution at ``low-LET''.

At a high LET, the reactive species are produced in large quantities and given an opportunity they would interact much faster than they diffuse and this would lead to their annihilation~\cite{Radicals}.
 However, at high values of LET there is another mechanism for transport of radicals, the collective flow of ion-induced shock waves. %The reason why this the mechanism is only called {\em possible} is that
The shock waves initiated by a large pressure difference and propagating radially from each ion's path were predicted in Ref.~\cite{prehydro} and discussed in a number of works within the MSA~\cite{MSAColl,CellSurSR16,natnuke,Vilnius,PabloSW,PabloRadDose,PabloPRA17,drop}; the transport of radicals with a collective flow including chemical reactions was investigated by means of molecular dynamics (MD) simulations in Ref.~\cite{PabloPRA17}.
%are still in the category of plausible effects since they have not yet been observed directly. In this paper, the reality of this effect is assumed.
As a result, the effective radii of ion tracks, within which the reactive species such as hydroxyl radicals and solvated electrons are transported, are substantially larger than those consistent with the diffusion transport mechanism. The observation of such large radii of ion tracks that can be inferred from the observation of ion tracks' interactions at large ion fluences is a strong argument in favor of the existence of collective flow. %It will be shown how to seamlessly match the transport of reactive species by a collective flow at high LET with a diffusive transport at low LET.

\subsection{Calculation of number of secondary electrons incident on a DNA target}

As has been shown in Ref.~\cite{Radicals}, the number densities of the first and second generations of secondary electrons are given by,
\begin{eqnarray}
n_1(t, r)=\frac{dN_1}{dx}\frac{1}{4\pi D_1 t}\exp\left(-\frac{r^2}{4 D_1 t}-\frac{t}{\tau_1} \right),\nonumber\\
n_2(t,r)=2\frac{1}{4\pi\tau_1} \frac{dN_1}{dx}\int_0^t \frac{1}{D_1 t'+D_2(t-t')} \nonumber \\ \times \exp\left(-\frac{r^2}{4( D_1 t'+D_2(t-t'))}-\frac{t-t'}{\tau_{2}}-\frac{t'}{\tau_1}\right)dt'.
\label{densities2}
\end{eqnarray}
%For the first generation, the problem can be solved analytically.
Since the characteristic spatial scale in radial direction is in nanometers and in the axial direction is micrometers, $\frac{dN_{1}}{dx}$ is assumed to be constant.

\begin{figure}
\begin{centering}
\resizebox{0.85\columnwidth}{!}
{\includegraphics{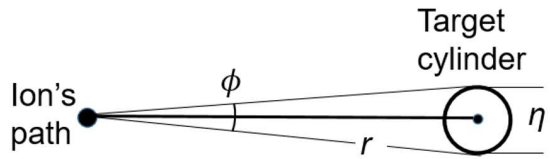}}
\caption{\label{fig.geom1} The geometry of the problem in the plane perpendicular to ion's path. The target cylinder that encloses a DNA twist is shown as a circle. Its diameter is $\eta$. The dimension $\xi$ is perpendicular to the plane of the figure.}
\end{centering}
\end{figure}

In this work, a target is chosen to be a rectangle of area $\xi\eta$, where $\xi=3.4$~nm and $\eta=2.3$~nm are the length of one twist and the diameter of a DNA molecule, respectively. Thus electrons or radicals hitting such a target would be hitting a rung of a DNA molecule masked by this target. The plane of the target is chosen to be parallel to the ion's path with dimension $\xi$ along and $\eta$ perpendicular to the path. This can be seen in Fig.~\ref{fig.geom1}. Then angle $\phi=2\arctan\frac{\eta/2}{r}$, where $r$ is the distance between the target and the path, inscribes the target in a plane perpendicular to the ion's path and $\xi\eta=r \phi \xi$.

\begin{figure}
\begin{centering}
\resizebox{0.85\columnwidth}{!}
{\includegraphics{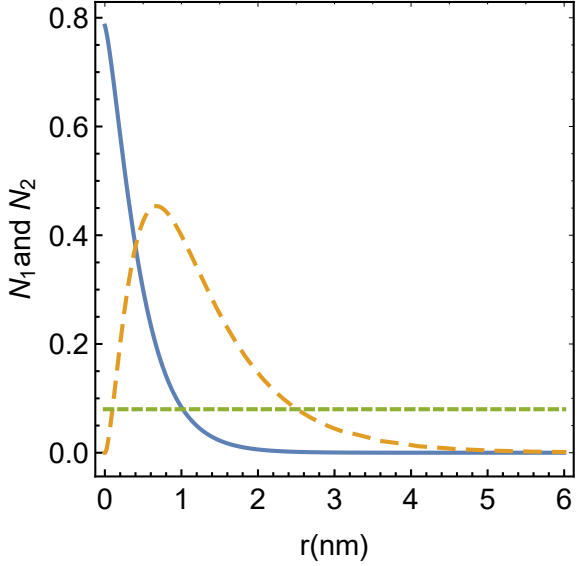}}
\caption{\label{fig.N1N2} The average numbers of simple lesions yield due to a single ion produced by secondary electrons of the first (solid line) and second (dashed line) generations, ${\cal N}_1(r)$ and ${\cal N}_2(r)$ as functions of distance from ion's path. These functions are calculated using (\ref{flux6}) and (\ref{flux7}) multiplied by $\Gamma_e=0.03$ (used in Ref.~\cite{MSAColl})  A straight (dotted) line is the similar value for reactive species, calculated using (\ref{radhighlet1}) with numbers used in Ref.~\cite{CellSurSR16}. The calculations are done for carbon ions near the Bragg peak. }
\end{centering}
\end{figure}

The number of first-generation electrons hitting the described target segment of area $r \phi \xi$ parallel to the ion's path per unit time is,
\begin{eqnarray}
\Phi_{1}=- \phi \xi r D_1 \frac{\partial n_1(r,t)}{\partial r}\nonumber \\
   = \frac{\phi}{2\pi}\frac{dN_{1}}{dx} \frac{r^2 \xi}{4 D_1 t^2}\exp\left(-\frac{r^2}{4 D_1 t}-\frac{t}{\tau_1}\right).
 \label{flux2}
\end{eqnarray}
Its integral over time,
\begin{eqnarray}
\int_0^\infty \Phi_1 dt=\frac{\phi}{2\pi}\int_0^\infty \frac{dN_{1}}{dx} \frac{r^2 \xi}{4 D_1 t^2}\exp\left(-\frac{r^2}{4 D_1 t}-\frac{t}{\tau_1}\right) \nonumber \\
=\frac{\phi}{2\pi}\frac{dN_{1}}{dx}\frac{r \xi}{\sqrt{D_1 \tau_1}}K_1\left(\frac{r}{\sqrt{ D_1\tau_1}}\right)~,
\label{flux5}
\end{eqnarray}
where $K_1$ is the Macdonald function (modified Bessel function of the second kind)~\cite{Abramowitz},
gives the total number of secondary electrons of the first generation that hit this area.
The second generation contribution is obtained similarly:
\begin{eqnarray}
\Phi_2(t,r)=-r \phi \xi D_2\frac{\partial n_2(r,t)}{\partial r}\nonumber \\
=\phi\frac{\xi r^2 D_2}{4\pi\tau_1} \frac{dN_1}{dx}\int_0^t \frac{1}{(D_1 t'+D_2(t-t'))^2} \nonumber \\ \times \exp\left(-\frac{r^2}{4( D_1 t'+D_2(t-t'))}-\frac{t-t'}{\tau_{2}}-\frac{t'}{\tau_1}\right)dt',
\label{flux6}
\end{eqnarray}
and then
\begin{eqnarray}
\int_0^\infty \Phi_2 dt=\phi\frac{\xi r^2 D_2}{4\pi\tau_1} \frac{dN_1}{dx}\int_0^\infty \int_0^t \frac{1}{(D_1 t'+D_2(t-t'))^2} \nonumber \\ \times \exp\left(-\frac{r^2}{4( D_1 t'+D_2(t-t'))}-\frac{t-t'}{\tau_{2}}-\frac{t'}{\tau_1}\right)dt'dt~~~
\label{flux7}
\end{eqnarray}
gives the number of secondary electrons of the second generation that hit the same area. The average number of simple lesions due to a single ion, ${\cal N}_e(r)$, can now be obtained as the sum,
\begin{eqnarray}
{\cal N}_e(r)={\cal N}_1(r)+{\cal N}_2(r)=\Gamma_e\int_0^\infty \Phi_1 dt+\Gamma_e\int_0^\infty \Phi_2 dt~,\nonumber\\
\label{flux8}
\end{eqnarray}
where $\Gamma_e$ is the probability for an electron to induce simple lesion such as a single strand break on a hit.
The values of ${\cal N}_1(r)$ and ${\cal N}_2(r)$ are shown in figure~\ref{fig.N1N2}.

Equation (\ref{flux8}) gives the average number of simple DNA lesions due to the secondary electrons of first and second generations as a function of distance of the target DNA segment fro the ion's path. The next step is to add to this the contribution of reactive species, $\Gamma_r N_r$. The latter depends on the value of LET, since at small values of LET the transport of radicals is defined by diffusion and at high values the collective flow is expected to dominate this process.

%Now, at each location we have $\Gamma_e N_e$ due to two generations of electrons that come from two different ions. We have to add $\Gamma_r N_r$ to include the radicals. Without the wave, we can do similar calculations assuming distributions obtained in Ref.~\cite{Radicals}. In order to include the wave, we can use the same idea as in Ref.~\cite{CellSurSR16} and use a step-function-like distribution of ${\cal N}_r=\Gamma_r N_r$ until the maximal radius, depending on the value of LET.

\subsection{Calculation of radical contribution for small values of LET}

The number of produced reactive species such as free radicals and solvated electrons depends on LET. If LET is not very high, it is expected that the number of reactive species is proportional to the secondary electron production, $dN_1/dx$, and, therefore, nearly linearly increases with the value of LET~\cite{MSAColl}. At sufficiently high values of LET, extra production of radicals is possible due to water radiolysis at locations adjacent to the ion's path. This effect has not yet been quantified and will be accounted for in future works along with the definition of the domain of LET, where this effect becomes significant. In this work, a linear dependence between the number of reactive species and LET is assumed and the difference between high and low values of LET is defined only by the mechanism of transport of the reactive species; at low LET, this transport is defined by diffusion. Moreover, this means that chemical reactions such as $2{\rm OH}\rightarrow {\rm H}_2{\rm O}_2$ and $e_{aq}+{\rm OH}\rightarrow {\rm OH}^-$ are rare and their frequency can be neglected compared to the diffusion term in the diffusion equation~\cite{Radicals}. Thus, the transport of reactive species in the low-LET case can be calculated, by solving a diffusion equation,
\begin{eqnarray}
\frac{\partial n_r}{\partial t}=D_r \nabla^2 n_r~,
\label{radlowlet1}
\end{eqnarray}
where $D_r$ is a diffusion coefficient for reactive species.

%Small values of LET in the context of this paper are defined in comparison with high values of LET at which the radial collective flow away from the ion's path due to the shock wave is the dominant way for propagation of reactive species. In Ref.~\cite{Radicals}, the initial distribution of reactive species was calculated regardless of the value of LET. It was then found that if the number densities of reactive species such as hydroxyl radicals or solvated electrons are high enough (at high LET), in absence of the shock wave they are expected to annihilate faster than they can spread at a significant distance. However, if the values of LET are small, the annihilation can be neglected and the main mechanism of propagation is diffusion.

The initial conditions for this equation can be taken from Ref.~\cite{Radicals},
\begin{eqnarray}
\frac{\partial n_{r}(r,t)}{\partial t}=\frac{dN_1}{d x}\delta^{(2)}(r)\delta(t)+\frac{n_{1}({r}, t)}{\tau_{1}}+\frac{n_{2}({r}, t)}{\tau_{2}},\nonumber \\
\label{oh.dist}
\end{eqnarray}
where the first term describes the species formed at sites of original ionizations by the projectile, while other two terms are due to inelastic processes involving secondary electrons of the first and secondary generations, respectively.
The production of reactive species, $n_{r}(r,t)$, through the mechanism of Eq.~(\ref{oh.dist}) ends by about 50~fs~\cite{Radicals}, by that time it is localized within 3~nm of the ion's path. These are the initial conditions for the following propagation of reactive species by the diffusion and/or collective flow, that happen on much larger scales, up to 100~ps in time and 50~nm in distance. Therefore, in this paper, a simplified initial condition,
\begin{eqnarray}
\frac{\partial n_{r}(r,t)}{\partial t}=K\frac{dN_1}{d x}\delta^{(2)}(r)\delta(t)~,
\label{oh.dist2}
\end{eqnarray}
where $K$ is the number of reactive species produced due to each secondary electron of the first generation ejected by an ion, will be used. The value of $K \approx 6$ can be evaluated as follows. The primary ionization produces H$_2$O$^+$, which is likely to produce a hydroxyl radical~\cite{hyd2}. The same happens when the secondary electron of the first generation ionizes another water molecule (and thus becomes an electron of second generation). Then two electrons of the second generation (the ionizing and ejected) can produce about four reactive species, two as a result of further energy loss in inelastic processes and two more if they become solvated electrons. A more accurate number for $K$ can be obtained if the probabilities of the above processes are combined and a $G$-value for an electron is calculated following a comprehensive radiochemical analysis.

The solution of Eq.~(\ref{radlowlet1}) with the initial condition (\ref{oh.dist2}) is given by,
\begin{eqnarray}
n_r(r,t) = K \frac{d N_1}{d x}\frac{1}{4\pi D_r t}\exp{\left(-\frac{r^2}{4D_r t}\right)}~.
\label{radlowlet3}
\end{eqnarray}
The next step is to find the number of reactive species incident on the target at a distance $r$ from the ion's path. We proceed similarly to (\ref{flux2}) and (\ref{flux5}).
\begin{eqnarray}
\Phi_{r}=- \phi \xi r D_r \frac{\partial n_r(r,t)}{\partial r}\nonumber \\
   = \frac{\phi}{2\pi}K \frac{dN_{1}}{dx} \frac{r^2 \xi}{4 D_r t^2}\exp\left(-\frac{r^2}{4 D_r t}\right)~,
 \label{radlowlet4}
\end{eqnarray}
and its integral over time is simply,
\begin{eqnarray}
\int_0^\infty \Phi_r dt=\frac{\phi \xi}{2\pi}K \frac{dN_{1}}{dx}~,
\label{radlowlet5}
\end{eqnarray}
which is independent of $r$. This seems unphysical, and a finite range must be introduced. Estimates for ranges can be obtained from the lifetimes of reactive species, such as solvated electrons and hydroxyl radicals, in liquid water, which are of the order of $10^{-4}$~s~\cite{Alpen,hyd2}. Then the ranges are estimated from diffusion ($\sqrt{D \tau}$), which gives about 300~nm with diffusion coefficients taken from Ref.~\cite{laverne89}.

\subsection{Calculation of radical contribution for large values of LET}

If the reactive species are formed in large quantities as a result of high-LET-ion's traverse, the collective flow due to the shock wave is the main instrument for the transport of these species away from the ion's path. Interestingly, ranges of propagation of radicals used to be in the realm of chemistry~\cite{hyd2,laverne89,Alpen}. However, in the case of high LET, this issue is addressed by physicists; the MD simulation with a use of MBN Explorer package~\cite{MBNX,MBNXCh} showed that the radius depends on the value of LET~\cite{PabloPRA17}, but more extensive investigation is needed to obtain a more detailed dependence.

In Ref.~\cite{CellSurSR16}, a simple model has been used to describe this transport. The value of average number of lesions at a distance $r$ from the ion's path, ${\cal N}_r=\Gamma_r N_r$, was considered to be a constant within a certain LET-dependent range $R$, i.e.,
\begin{eqnarray}
{\cal N}_r(r)={\cal N}_{r}\Theta(R-r)~,
\label{radhighlet1}
\end{eqnarray}
where $\Theta$ is a Heaviside step function. Value ${\cal N}_r$ also depends on the degree of oxygenation of the medium, since the concentration of oxygen resolved in the medium affects the number of formed radicals as well as the effectiveness of lesion repair. In principle, more information about ${\cal N}_r$ is needed. For example, at high LET, more reactive species are expected to be produced through the radiolysis of water in the cores of ion tracks at times $\ge~50$~fs after the energy transfer from secondary electrons to the medium has taken place. This process can now be studied by molecular dynamics (MD) simulations using the MBN Explorer package~\cite{MBNX,MBNXCh}, which is capable of resolving corresponding temporal and spatial scales.

The comprehensive picture of transport of reactive species includes the diffusion (dominant at low values of LET), collective flow (dominant at high values of LET), and chemical reactions. With this understanding, as LET increasing Eq.~(\ref{radlowlet5}) should gradually transform into (\ref{radhighlet1}). Hopefully, this will be done in a future work.

\subsection{Calculation of production of DNA lesions per unit length of ion propagation in the case of one and more ions}

Within the MSA, the probability of lesions is calculated using a Poisson statistics and the next step is the calculation of the average value of simple lesions, ${\cal N}$.
%, before they can be combined into lethal lesions and integrated.
%%%%%%%%%%%%%%%%%%%%%%%%%%%%%%%%%%%%%%%%%%%%%%%%%%%%%%%%%%%%%%%%%%%5
First, we consider a case when ion tracks do not interfere, which is realized when the doses are not large enough. Then the effect of a single ion is just multiplied by the number of ions passing through the target.
In this case, this value is calculated as,
\begin{eqnarray}
{\cal N}={\cal N}_1(r)+{\cal N}_2(r)+{\cal N}_r(r)~.
\label{n1}
\end{eqnarray}
Based on this, the probability of lethal damage according to the criterion of lethality~\cite{MSAColl,CellSurSR16},
\begin{eqnarray}
P_l(r)=\lambda \sum_{\nu=3}^\infty\frac{{\cal N}^\nu}{\nu !}\exp{\left[-{\cal N}\right]}~,
\label{nrob.comp}
\end{eqnarray}
where $\lambda=0.15$. This criterion states that three DNA lesions one of which is a double strand break have to occur within two DNA twists. The probability given by Eq.~(\ref{nrob.comp}) is then integrated over space ($2\pi rdr$) giving the number of lethal lesions per unit segment of ion's path, $dN_l/dx$,
\begin{eqnarray}
\frac{d N_{l1}}{d x}=2\pi n_s \int_0^\infty P_l(r)r dr~.
\label{nlb0}
\end{eqnarray}
This quantity was calculated in Ref.~\cite{MSAColl}, but in this paper we are going to consider the interference of two or more ions and calculate a similar quantity in these cases.

\begin{figure}
\begin{centering}
\resizebox{0.85\columnwidth}{!}
{\includegraphics{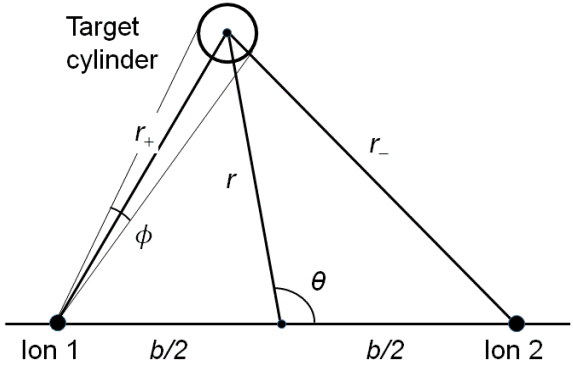}}
\caption{\label{fig.geom} The geometry of the problem in the plane perpendicular to ions' paths, separated by distance $b$. The target cylinder that encloses a DNA twist is shown as a circle. Its diameter is $\eta$. The dimension $\xi$ is perpendicular to the plane of the figure. The geometry for a single ion is corresponds to $b=0$.}
\end{centering}
\end{figure}

\begin{figure}
\begin{centering}
\resizebox{0.85\columnwidth}{!}
{\includegraphics{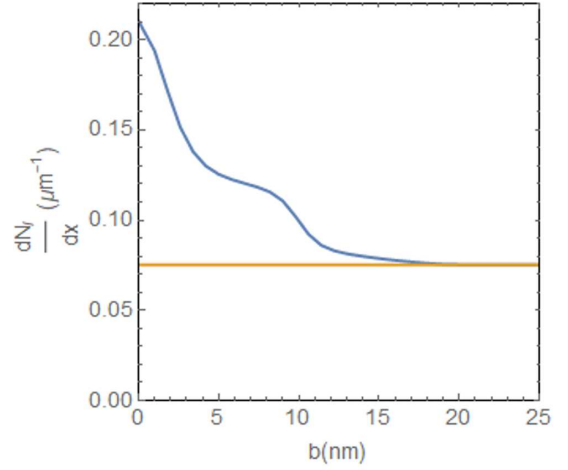}}
\caption{\label{fig.Nlb} The average number of lethal lesions produced by two ions whose parallel paths are separated by distance $b$ calculated using (\ref{nlb}) as a function of distance $b$. A straight line shows the doubled number of lesions produced by a single ion.}
\end{centering}
\end{figure}

Now, let us consider two ions whose paths are parallel to each other and separated by a distance $b$ and calculate the number of lethal lesions per unit segment of ions' paths, $dN_{l2}/dx$, which includes their interference. Figure~\ref{fig.geom}
illustrates the geometry of this problem. Similarly to Eq.~(\ref{n1}), the calculation starts with ${\cal N}$, which now also depends on angle $\theta$,
\begin{eqnarray}
{\cal N}={\cal N}_1(r, \theta)+{\cal N}_2(r, \theta)+{\cal N}_r(r, \theta).
\label{assemb1}
\end{eqnarray}
%and the spatial integration of $r$- and $\theta$-dependent (\ref{nrob.comp}) becomes more difficult.
Each term in (\ref{assemb1}) is obtained according to the sum,
\begin{eqnarray}
{\cal N}_i(r, \theta)={\cal N}_i(r_+)+{\cal N}_i(r_-)~,
\label{assemb2}
\end{eqnarray}
where
\begin{eqnarray}
r_+=\sqrt{r^2+b^2/4+r b \cos \theta}~, \nonumber \\
r_-=\sqrt{r^2+b^2/4-r b \cos \theta})~.
\label{assemb3}
\end{eqnarray}
%Figure~\ref{fig.geom} clarifies the geometry of this arrangement.
%In particular, the value of ${\cal N}_r(r,\theta)$ is given by
%\begin{eqnarray}
%{\cal N}_r(r,\theta)
%={\cal N}_{r}\left[\Theta(R-r_+)+\Theta(R-r_-)\right]~.
%\end{eqnarray}
After ${\cal N}(r, \theta)$ is assembled, it is inserted in (\ref{nrob.comp}) and the latter is integrated:
\begin{eqnarray}
\frac{d N_{l2}}{d x}(b)=4 n_s \int_0^\infty\int_0^{\pi/2}P_l(r,\theta)r dr d\theta~,
\label{nlb}
\end{eqnarray}
where $n_s$ is the target density calculated as in Ref.~\cite{CellSurSR16}. The result of this integration,
$\frac{d N_{l2}}{d x}(b)$, is the number of lethal lesions done by a pair of ions whose parallel paths are at a distance $b$ from each other, is shown in figure~\ref{fig.Nlb} as a function of $b$.
It is compared to a doubled $\frac{d N_{l1}}{d x}$, calculated by (\ref{nlb0}). When $b$ becomes large enough, the interference vanishes and these quantities are equal.

The effect of the contribution of track interference apparent in Fig.~\ref{fig.Nlb} is considerably reduced because the yield $\frac{d N_{l2}}{d x}$ has to be multiplied by the probability of paths being within distance $b$ apart, calculated in the next section. In order for the interference effect in $\frac{d N_{l2}}{d x}$ to be substantial, distance $b$ has to be sufficiently small. As a consequence large doses and, therefore, ion fluences are required to maintain this substantiality. The contribution of $\frac{d N_{l2}}{d x}$ above $2\frac{d N_{l1}}{d x}$ is the first correction to the tracks interference effect, since $\frac{d N_{l3}}{d x}$ and higher order terms may contribute as well; this will be done below.

%However, there are two reasons why these contributions are not calculated in this paper. First, as will be shown below, the interference does require high doses and even higher doses are needed in order to appreciate the importance of $\frac{d N_{l3}}{d x}$ term. Second, the arbitrary parameter $b$ is introduced in a unique way as a distance between the paths and it can be chosen so that the effect of interference is maximized. Additional assumptions are needed in order to calculate $\frac{d N_{l3}}{d x}$, e.g., that three paths are equidistant from each other, or some other configuration that would maximize that effect. Such assumptions would require additional mathematical analysis and lead the study away from the original purpose of exploration of effect of ion tracks interference at high doses on the formation of shoulders in cell-survival curves.
%This is similar to a number due to a single ion, calculated in Ref.~\cite{MSAColl},
%\begin{eqnarray}
%\frac{d N_{l1}}{d x}=2\pi n_s \int_0^\infty P_l(r)r dr~,
%\label{nlb0}
%\end{eqnarray}
%where $P_l(r)$ is still obtained from (\ref{nrob.comp}), but with ${\cal N}$ calculated from (\ref{n1}).
%This number (multiplied by two), independent of $b$, is shown in figure~\ref{fig.Nlb}. Certainly the results coincide as soon as tracks' overlap is over at a certain value of $b$.

\subsection{Calculation of the probability of paths overlap}

Let us consider a monochromatic ion beam with a uniform fluence $F$ (far from the edges of the beam) defined as the number of ions incident on a unit area of the target. Let $S_e=-\frac{dE}{dx}$, where $E$ is the ion energy and $x$ is the longitudinal coordinate, be the electronic stopping power of these ions at a given position $x$. Then the average dose deposited in a slice perpendicular to the ion beam between $x$ and $x+dx$ is given by
\begin{eqnarray}
d=F S_e~.
\label{dose}
\end{eqnarray}
Then, and the average cross sectional area per ion is,
\begin{eqnarray}
A_1=F^{-1}=\frac{S_e}{d}~.
\label{dose2}
\end{eqnarray}
The probability for paths of two ions to be within a distance $b$ can be found using a Poisson statistics. From Eq.~(\ref{dose2}), the average number of ion paths crossing area $A$ is $N_A=FA$. If we choose the $x$-axis as a path of a given ion, the the probability that another ion's path is within area $A=\pi b^2$ is,
\begin{eqnarray}
{\cal P}(b)=\frac{1}{A_1}\int_0^A N_A e^{-N_A}dA \nonumber \\ =F^2\int_0^b \pi r^2 e^{-F\pi r^2}2\pi r dr \nonumber \\
=1 - e^{-F \pi b^2} (1 + F \pi b^2)
\label{dose3}
\end{eqnarray}
This probability is normalized,
\begin{eqnarray}
F^2\int_0^\infty \pi r^2 e^{-F\pi r^2}2\pi r dr=1~,
\label{dose3n}
\end{eqnarray}
and has a maximum when $AF=1$, i.e., at $A=A_1$, corresponding to (\ref{dose2}).
Thus, the probability that another ion is propagating parallel to the $x$-axis within distance $b$ from the first ion's path is given by Eq.~(\ref{dose3}), which after substitution of (\ref{dose}) becomes,
\begin{eqnarray}
{\cal P}(b,d)=1 - e^{-\frac{\pi b^2}{S_e}d} (1 + \frac{\pi b^2}{S_e}d)~.
\label{pro3}
\end{eqnarray}
This is the (dose-dependent) probability that the paths of two ions are within distance $b$. Apparently, ${\cal P}(b\rightarrow\infty,d>0)={\cal P}(b>0,d\rightarrow \infty)=1$.

\subsection{Calculation of the yield of lethal lesions}

At this point, the results of previous sections can be combined in the expression for the yield of lethal lesions. Such an expression was obtained in Refs.~\cite{MSAColl,CellSurSR16} for the case of non-interfering ion paths as,
\begin{eqnarray}
Y_{1}
= \frac{{d}N_{l1}}{{d}x} \, \bar{z} \, N_{\rm ion}(d)~,
\label{eq08a}
\end{eqnarray}
where $N_{\rm ion}$ is the number of ions that traverse a target, and $\bar{z}$ is the average length of trajectory of ion's traverse. This yield is a product of the yield per unit length of ion's path and the average length within a target passed by all ions (${\bar z}N_{ion}$). In order to derive an expression accounting for the interference between the ion tracks, the yield per unit path length due to a pair of ions~(\ref{nlb}) has to substitute $\frac{{d}N_{l1}}{{d}x}$ in Eq.~(\ref{eq08a}). For a single pair of ions at a given distance $b$,
\begin{eqnarray}
Y_{2}(b)
= \frac{1}{2}\frac{{d}N_{l2}}{{d}x}(b) \, \bar{z} \, N_{\rm ion}(d){\cal P}(b,d)~,
\label{eq08aa}
\end{eqnarray}
where the factor $1/2$ is inserted because $\frac{{d}N_{l2}}{{d}x}$ includes two ions. ${\cal P}(b,d)$ is a probability that these ions are within distance $b$ apart. It is easy to see that in the limit of $b \rightarrow \infty$ Eq.~(\ref{eq08aa}) becomes identical to Eq.~(\ref{eq08a}). Equation~(\ref{eq08aa}) can be generalized for average number of ions traversing target as
\begin{eqnarray}
Y_{2}
= \frac{1}{2} \, \bar{z} \,N_{\rm ion}(d)\, \frac{\sum_{i\neq j} \frac{{d}N_{l2}}{{d}x}(b_{ij}){\cal P}(b_{ij},d)}{N_{\rm ion}(N_{\rm ion}-1)}\nonumber \\
= \frac{1}{2} \, \bar{z} \, N_{\rm ion}(d) \frac{{d}N_{l2}}{{d}x}({\bar b}){\cal P}({\bar b},d)~.
\label{eq08aaa}
\end{eqnarray}
In this generalization,  $N_{\rm ion}(N_{\rm ion}-1)$ terms are included in the sum and all interference effects are taken into account and then some average ${\bar b}$ is introduced. This parameter, ${\bar b}$, is chosen below in such a way that the interference effect is maximized. However, Eq.~(\ref{eq08aaa}) does not give us the final result, since it assumes the {\em pairing} of ions and does not allow them to be independent; e.g., at small doses, ${\cal P}({\bar b},d)$ can be so small that Eq.~(\ref{eq08aaa}) will predict a result much less than Eq.~(\ref{eq08a}), which is not physical. The correct expression is
\begin{eqnarray}
Y_{12}
= \frac{{d}N_{l1}}{{d}x} \, \bar{z} \, N_{\rm ion}(d)(1-{\cal P}({\bar b},d))\nonumber \\ +\frac{1}{2} \, \bar{z} \, N_{\rm ion}(d) \frac{{d}N_{l2}}{{d}x}({\bar b}){\cal P}({\bar b},d)\,,
\label{eq08b}
\end{eqnarray}
which is a mathematical expectation of the yield due to non-interfering and interfering ions. This equations suggests the way to further generalize this expression to include the interference of more than two ion tracks. This equation can be written as,
\begin{eqnarray}
Y_{l}
= \bar{z} \, N_{\rm ion}(d)\sum_{n=1}^{\infty} \frac{1}{n}\, \frac{{d}N_{ln}}{{d}x} {\cal P}_n({\bar b_n},d)\,,
\label{eq08bb}
\end{eqnarray}
where index $n$ indicates how many tracks are involved in the overlap and ${\cal P}_1=1-\sum_{n=2}^{\infty}{\cal P}_n$. Apparently, Eq.~(\ref{eq08bb}) gives Eq.~(\ref{eq08b}) if the sum is truncated after $n=2$. However, in order to finish this problem, the yields $\frac{{d}N_{ln}}{{d}x}$ for $n>2$ have to be calculated. Since the problem of inclusion of orders higher than two is marginal to this paper, this is done in the Appendix.

After accounting for the interaction between ion tracks, the expression for survival probability of Refs.~\cite{MSAColl,CellSurSR16,sobp},
\begin{equation}
\ln{\Pi_{surv}}=-Y_{l}
\label{logprob.surv}
\end{equation}
remains the same, but the dose dependence in Eqs.~(\ref{eq08aa})-(\ref{eq08bb}) is non-linear and the survival curve becomes shouldered regardless of the repair mechanisms.

%Interestingly, from the mathematical point of view, at large doses this dependence becomes linear and then a question about including another ion's track is due. This may not be the most practical question at this point, but it may need to be answered if biological experiments drive in that direction.

\begin{figure}
\begin{centering}
\resizebox{0.9\columnwidth}{!}
{\includegraphics{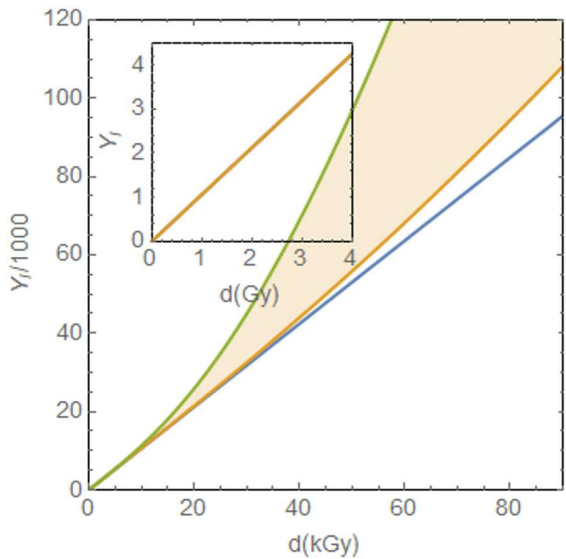}}
\caption{\label{fig.Yd} The yield of lethal lesions in A549 cells produced by carbon ions with LET of 120~eV/nm as a function of dose. At very high values of dose the yield calculated with account for interference of ion tracks (\ref{eq08b}) varies from that without tracks interference (\ref{eq08a}); the shouldered orange curve deviates from the straight line. In both calculations, the range of propagation of reactive species $R$ is taken to be 10~nm, as was done in Refs.~\cite{MSAColl,CellSurSR16}. The green line corresponds to the same yield, calculated with $R=30$~nm. The shaded region shows the range of change of the yield the value of $R$ varies between 10 and 30 nm. The inset shows the same dependencies at clinically reasonable doses; the fact that they coincide explains why for 120~eV/nm carbon ions the survival curve is a straight line and can exhibit a shoulder only due to repair effects.  }
\end{centering}
\end{figure}

The dependence of yield $Y_{l}$ calculated using Eq.~(\ref{eq08b}) on dose is shown in figure~\ref{fig.Yd} (orange solid line); it is compared to the straight line dependence obtained from Eq.~(\ref{eq08a}). These calculations were done for A549 cells irradiated with carbon ions at LET of 120~eV/nm.
The calculations are done for a particular value of $b=8$~nm, which is the value of $b$ at which the overlap effect is maximized.
The kGy scale of this figure as well as the average number of DNA lesions of $\sim 5\times 10^4$ explains why the question of path interference was not raised earlier in Refs.~\cite{MSAColl,CellSurSR16,sobp}; this effect may not be relevant for medical applications related to cell killing with carbon ions. The inset showing the coincidence of the same dependencies at clinically reasonable doses further corroborates this point.

The above calculations were done with the range of reactive species propagation $R=10$~nm; this is the value that was used for carbon ions in the Bragg peak region in Refs.~\cite{MSAColl,CellSurSR16}. However, at this point this number is not known; an MD study of the dependence of $R$ on LET is in progress. However, in the mean time, an inverse problem can be addressed: starting from the experimental results, find the range of propagation of reactive species. Such a study can lead to a better understanding of transport of reactive species and the radiation damage scenario in general. A recent investigation of this problem~\cite{Kowalska17} using the local effect model (LEM)~\cite{LEM96} shows the interest of experimentalists to this issue, and this paper is providing a different tool for these studies, based on the MSA. A (green) line shown in figure~\ref{fig.Yd} indicates the same dependence of yield of lesions for carbon ions calculated with $R=30$~nm instead of 10 in the dependence shown with the orange curve; the shaded area indicates the span of results for $R$ between 10 and 30~nm. As expected, the larger the range, the earlier is the departure of a shouldered curve from the straight line.

\begin{figure}
\begin{centering}
\resizebox{0.9\columnwidth}{!}
{\includegraphics{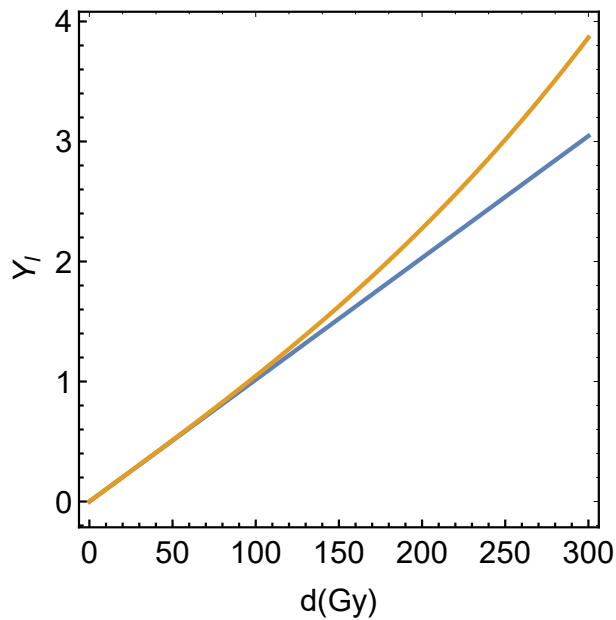}}
\caption{\label{fig.Yd2} The yield of lethal lesions in A549 cells produced by protons ions with LET of 10~eV/nm as a function of dose. The survival curve becomes shouldered at a much smaller value of dose compared to that with carbon ions because the fluence increases inversely proportionally to LET.}
\end{centering}
\end{figure}

Another example of track interference is shown in figure~\ref{fig.Yd2}, where the number of lesions vs. dose is shown for protons instead of carbon ions (with $R=10$~nm). The deviation of the curve accounting for track interference from a straight line occurs at a much smaller dose.
This happens because the LET is about $12$ times smaller: the dose is reduced by this factor from Eq.~(\ref{pro3}) and even more significantly by the reduction of the number of ionization events $dN_1/dx$ and $d N_{ln}/d x$ as a consequence. In general, the requirement of high doses at lower values of LET leads to the requirement of large ion fluences and ion track interference.
We predict that the interference of tracks will be important in plasmid DNA experiments as well as in experiments with laser-driven proton beams, where much larger ion fluences are inevitable.

\section{Conclusion}

This methodological work presents an approach to the calculation of survival curves for targets irradiated with ions in cases where ion fluences are large enough so that the track interference becomes significant. The calculations are done using the multiscale approach to the physics of radiation damage with ions and one can see the particular effects that bring about the shouldered dependence of damage on the dose. This method can be applied to many experiments on cells and plasmid DNA conducted around the world in order to understand more about the transport of reactive species in ion tracks and the effectiveness of enzymatic repair vs. track interaction effects. This method can also become vital for assessment of radiation damage done by laser-driven proton beams.

\begin{appendix}
\section*{Appendix: Calculation of $dN_{ln}/dx$ and ${\cal P}_n$ for $n>2$}

The logic of calculation of higher orders of interference of ion tracks is similar to that used above for the calculation for $n=1$ and $n=2$. The corresponding ${\cal N}$ is obtained from an equation similar to (\ref{n1}) and (\ref{assemb1}), which is then substituted into Eq.(\ref{nrob.comp}), and integrated as Eqs.~(\ref{nlb0}) and (\ref{nlb}).

Geometrical arrangements of ion paths for $n>2$ are not unique. Let us consider a plane perpendicular to the direction of propagation of ions. In this plane the paths appear as dots. We suggest arranging these dots at vertices of equilateral polygons starting with $n=3$. Such arrangements minimize the number of parameters introduced. Indeed, only one parameter, e.g., a distance between the center and a vertex $b_n$ is sufficient for each $n$.

Then ${\cal N}(r,\theta)$ is combined similarly to (\ref{assemb1}). Only, instead of (\ref{assemb3}), there are more complicated expressions such as,
\begin{eqnarray}
r_1=\sqrt{r^2+b_3^2-2r b_3 \cos \theta}~, \nonumber \\
r_2=\sqrt{r^2+b_3^2-2r b_3 \cos (2\pi/3 - \theta)}~, \nonumber \\
r_3=\sqrt{r^2+b_3^2-2r b_3 \cos (4\pi/3 - \theta)})~,
\label{app1}
\end{eqnarray}
for the case of $n=3$, which can be generalized as,
\begin{eqnarray}
r_{n1}=\sqrt{r^2+b_n^2-2r b_n \cos \theta}~, \nonumber \\
r_{n2}=\sqrt{r^2+b_n^2-2r b_n \cos (2\pi/n - \theta)}~, \nonumber \\
...\nonumber \\
r_{nk}=\sqrt{r^2+b_n^2-2r b_n \cos (2\pi k/n - \theta)}~,\nonumber \\
r_{nn}=\sqrt{r^2+b_n^2-2r b_n \cos (2\pi (n-1)/n - \theta)}~,
\label{app2}
\end{eqnarray}
where $1<k<n$, for any $n>2$.

Probability ${\cal P}_n$ can be calculated similarly to ${\cal P}_2$ starting with equation similar to~(\ref{dose3}):
\begin{eqnarray}
\frac{1}{(n-1)!}\frac{1}{A_1}\int_0^A N_A^{n-1} e^{-N_A}d A\nonumber \\=\frac{2 F^n \pi^n}{(n-1)!}\int_0^{b_n}r^{2n-1} e^{-F\pi r^2} dr\nonumber \\
=1-\frac{\Gamma(n,N_A)}{(n-1)!}~,
\label{dose3app}
\end{eqnarray}
where $\Gamma(n,N_A)=\int_{N_A}^{\infty} t^{n-1}e^{-t} dt$ is an incomplete $\Gamma$-fun\-ction.
 Then the probability that $n$ ions are propagating parallel to the $x$-axis within radius $b_n$ is,
\begin{eqnarray}
{\cal P}_n=\frac{2 F^n \pi^n}{(n-1)!}\int_0^{b_n} r^{2n-1} e^{-F\pi r^2}dr~.
\label{pro1app}
\end{eqnarray}

\end{appendix}

\bibliography{bibliography1}

\end{document}